\documentclass[11pt,a4paper]{article}
\pdfoutput=1
\usepackage{a4wide}
\usepackage{amsmath}
\usepackage{amssymb}
\usepackage{amsfonts}
\usepackage{cite}
\usepackage{color}
\usepackage{adjustbox}
\usepackage{multirow}
\usepackage{pdflscape}
\usepackage{gensymb}
\usepackage{graphicx}
\usepackage{diagbox}
\usepackage{pifont}
\usepackage{bigstrut}
\usepackage[small,bf]{caption}
\usepackage{tabulary}
\usepackage{booktabs}
\usepackage{stackengine}
\usepackage{hyperref}

\def\1{\mathbf{1}}
\def\3{\mathbf{3}}
\def\2{\mathbf{2}}

\def\D{\Delta}

\allowdisplaybreaks
\makeatletter
\g@addto@macro\bfseries{\boldmath}
\makeatother

\begin{document}

\begin{titlepage}
\renewcommand*{\thefootnote}{\fnsymbol{footnote}}

\begin{center}
{\bf\Large Viability of $A_4$, $S_4$ and $A_5$ Flavour Symmetries}\\[2mm]
{\bf\Large in Light of the First JUNO Result}\\[8mm]
S.~T.~Petcov$^{\,a,b,}$\footnote{Also at Institute of Nuclear Research and Nuclear Energy, Bulgarian Academy of Sciences, 1784 Sofia, Bulgaria.} 
and
A.~V.~Titov$^{\,c,d,}$\footnote{\href{mailto:arsenii.titov@unipd.it}{\texttt{arsenii.titov@unipd.it}}} \\
\vspace{8mm}
$^{a}$\,{\it INFN/SISSA, Via Bonomea 265, 34136 Trieste, Italy} \\
\vspace{2mm}
$^{b}$\,{\it Kavli IPMU (WPI), UTIAS, University of Tokyo, \\
5-1-5 Kashiwanoha, 277-8583 Kashiwa, Japan} \\
\vspace{2mm}
$^{c}$\,{\it Dipartimento di Fisica e Astronomia ``Galileo~Galilei'', 
Università degli Studi di Padova,  
Via Francesco~Marzolo 8, 35131 Padova, Italy} \\
\vspace{2mm}
$^{d}$\,{\it INFN, Sezione di Padova,  
Via Francesco~Marzolo 8, 35131 Padova, Italy}
\end{center}
\vspace{8mm}

\begin{abstract}
\noindent 
We update the analysis of the viability 
of the lepton mixing patterns originating 
from $A_4$,~$S_4$ and $A_5$ discrete flavour symmetries
and leading to predictions for the solar 
neutrino mixing angle, $\theta_{12}$.
We perform a statistical analysis using as an input 
(i) the results of the latest global fit to neutrino oscillation data, 
and (ii) the first JUNO measurement of~$\sin^2\theta_{12}$.
Out of the five (four) cases compatible with the global data at $3\sigma$ 
for normal (inverted) neutrino mass ordering, 
only three (two) cases 
remain compatible with the global data at the same confidence level 
after taking into account the JUNO result.
\end{abstract}
\end{titlepage}
\setcounter{footnote}{0}

\paragraph{Introduction.}
After only 59.1 days of data-taking, the JUNO experiment~
\cite{JUNO:2015zny,JUNO:2015sjr,JUNO:2021vlw,JUNO:2025fpc}, 
conceived in~\cite{Petcov:2001sy,Choubey:2003qx,Learned:2006wy,Li:2013zyd},
released its first physics result, 
determining the solar 
neutrino mixing angle, $\theta_{12}$, 
or more precisely $\sin^2\theta_{12}$, and the solar 
neutrino mass-squared difference, $\Delta m^2_{21} \equiv m_2^2 - m_1^2$, 
to unprecedented precision~\cite{JUNO:2025gmd}:
\begin{equation}
\sin^2\theta_{12} = 0.3092 \pm 0.0087 
\qquad \text{and} \qquad
\Delta m^2_{21} = (7.50 \pm 0.12) \times 10^{-5}~\text{eV}^2\,,
\label{eq:JUNO}
\end{equation}
reducing the relative $1\sigma$ uncertainty in the determination of 
$\sin^2\theta_{12}$ to 2.81\% and that in the determination of 
$\Delta m^2_{21}$ to 1.55\%. This result improves the precision by a factor 
of 1.6 with respect to the previous measurements of $\sin^2\theta_{12}$ 
performed by Super-Kamiokande+SNO~\cite{Super-Kamiokande:2023jbt}, 
and of $\Delta m^2_{21}$ performed by KamLAND~\cite{KamLAND:2013rgu}.

In this \textit{letter}, we show that the JUNO measurement has strong 
implications for the lepton mixing patterns originating from 
non-Abelian discrete flavour symmetries 
(see~\cite{Feruglio:2019ybq,Petcov:2017ggy,King:2013eh,Ishimori:2010au,Altarelli:2010gt} for reviews) and yielding sharp predictions for $\sin^2\theta_{12}$. 
Our focus is on the small (in terms of the number of elements) groups 
$A_4$, $S_4$ and $A_5$ featuring a 3-dimensional irreducible representation. 
The underlying assumption is that a flavour symmetry group effective 
at some high-energy scale is broken down at low energies in such 
a way that the charged lepton and neutrino mass matrices preserve certain 
residual symmetries described by its Abelian subgroups. 
A comprehensive analysis of such mixing patterns originating from all possible 
residual symmetries has been performed in~\cite{Girardi:2015rwa}, 
and their compatibility with the global neutrino oscillation data available 
in the beginning of 2018 has been assessed in~\cite{Petcov:2018snn}.

In this work, we first update our previous analysis~\cite{Petcov:2018snn} 
employing the results of the latest global fit to neutrino oscillation 
data performed by the NuFIT collaboration in 
September 2024~\cite{Esteban:2024eli,NuFITv6Sep2024},
see Table~\ref{tab:NuFIT6.0}.%
\footnote{See~\cite{Capozzi:2025wyn} for an independent global analysis 
of neutrino oscillation data that yields similar results.}
\begin{table}[t]
\small
\centering
\renewcommand*{\arraystretch}{1.5}
\begin{tabular}{|l|cc|cc|}
\hline
& \multicolumn{2}{c|}{Normal ordering (NO)} & \multicolumn{2}{c|}{Inverted ordering (IO)} \\
\cline{2-5}
Parameter & Best-fit value $\pm 1\sigma$ & $3\sigma$ range & Best-fit value $\pm 1\sigma$ & $3\sigma$ range \\
\hline
$\sin^2\theta_{12}$ & $0.308^{+0.012}_{-0.011}$ & $0.275 \to 0.345$ & $0.308^{+0.012}_{-0.011}$ & $0.275 \to 0.345$ \\
$\sin^2\theta_{23}$ & $0.470^{+0.017}_{-0.013}$ & $0.435 \to 0.585$ & $0.550^{+0.012}_{-0.015}$ & $0.440 \to 0.584$ \\
$\sin^2\theta_{13}$ & $0.02215^{+0.00056}_{-0.00058}$ & $0.02030 \to 0.02388$ & $0.02231^{+0.00056}_{-0.00056}$ & $0.02060 \to 0.02409$ \\
$\delta~[^\circ]$ & $212^{+26}_{-41}$ & $124 \to 364$ & $274^{+22}_{-25}$ & $201 \to 335$ \\
\hline
\end{tabular}
\caption{Best-fit values along with $1\sigma$ uncertainties and $3\sigma$ ranges of the lepton mixing parameters obtained in the latest global analysis of neutrino oscillation data (including Super-Kamiokande) 
performed by the NuFIT collaboration in September 2024~\cite{Esteban:2024eli,NuFITv6Sep2024}.}
\label{tab:NuFIT6.0}
\end{table}
We find that five (four) cases leading to sharp predictions 
for $\sin^2\theta_{12}$ 
are compatible with the global data at the $3\sigma$ confidence level, 
assuming normal (inverted) neutrino mass ordering, 
denoted further as NO (IO). 
Next, we perform a statistical analysis including the recent JUNO 
result, eq.~\eqref{eq:JUNO}. 
We find that only three (two) mixing patterns survive at 
$3\sigma$ after the JUNO measurement. 
We also discuss the predictions of these mixing patterns for 
the atmospheric neutrino mixing angle, $\theta_{23}$, and the 
Dirac CP-violating (CPV) 
phase, $\delta$, present in the PMNS lepton mixing matrix.

We note that beyond corroborating/falsifying lepton mixing patterns~\cite{Zhang:2025jnn,He:2025idv}, 
potentially based on flavour and/or CP symmetries~\cite{Ge:2025csr,Chen:2025afg,Jiang:2025hvq}, 
the first JUNO result has implications for 
neutrinoless double beta decay~\cite{Ge:2025cky},
probing unitarity (violation) of the lepton mixing matrix~\cite{Huang:2025znh}, 
flavour composition of astrophysical neutrinos~\cite{Xing:2025xte},
and (dark) matter effects on neutrino oscillations~\cite{Chao:2025sao,Li:2025hye}.

This \textit{letter} is organised as follows. First, we briefly summarise 
the symmetry-based mixing patterns of interest. 
Then, after describing the procedure used in our statistical analysis, 
we present our results in terms of likelihood plots for $\sin^2\theta_{12}$, 
$\cos\delta$ and $\sin^2\theta_{23}$. 
Finally, we summarise our findings and draw conclusions.

\paragraph{Mixing patterns predicting $\sin^2\theta_{12}$.}
Residual  symmetries $G_e$ and $G_\nu$ of the charged lepton 
and neutrino mass matrices, respectively, 
lead to relations between 
(i) (some of) the mixing angles and 
(ii) (some of) the mixing angles and the 
Dirac CPV phase, $\delta$, in the PMNS matrix.
As shown in~\cite{Girardi:2015rwa} (see also~\cite{Petcov:2018snn}), 
if $G_e = Z_k$, $k > 2$ or $Z_m \times Z_n$, $m,n \geq 2$ 
and $G_{\nu} = Z_2$, the solar neutrino mixing parameter 
$\sin^2\theta_{12}$ (cosine of the CPV phase, $\cos\delta$) is expressed 
in terms of $\theta_{13}$ ($\theta_{13}$ and $\theta_{23}$) 
and the parameters $\theta^\circ_{ij}$ fixed by the underlying symmetries.
This pattern of residual symmetries, denoted as pattern B, 
leads to two cases, B1 and B2, 
depending on the plane, not fixed by 
the $G_\nu = Z_2$ symmetry, 
in which a $U(2)$ transformation  acts. 
In case B1, we have
\begin{equation}
\sin^2 \theta_{12} = \frac{\sin^2 \theta^{\circ}_{12}}{1 - \sin^2 \theta_{13}}\,, \\
\label{eq:ss12B1}
\end{equation}
\begin{equation}
\cos\delta = -\frac{\cos^2 \theta_{13} (\cos^2 \theta^{\circ}_{12} \cos^2 \theta^{\circ}_{23} - \cos^2 \theta_{23}) + \sin^2 \theta^{\circ}_{12} (\cos^2 \theta_{23} - \sin^2 \theta_{13} \sin^2 \theta_{23})}
{\sin 2 \theta_{23} \sin \theta_{13} |\sin \theta^{\circ}_{12}| (\cos^2 \theta_{13} - \sin^2 \theta^{\circ}_{12})^{\frac{1}{2}}}\,,
\label{eq:cosdeltaB1}
\end{equation}
where $\sin^2\theta_{13}$ and $\sin^2\theta_{23}$ are functions of two 
\textit{a priori} free real parameters~---~an angle $\theta \in [0,\pi)$ 
and a phase $\phi \in [0,2\pi)$~---~entering the $U(2)$ transformation~\cite{Girardi:2015rwa} (see also \cite{Blennow:2020snb}):
\begin{equation}
\sin^2 \theta_{13} = \cos^2 \theta^{\circ}_{12} \sin^2 \theta \,, 
\label{eq:ss13B1}
\end{equation}
\begin{equation}
\sin^2 \theta_{23} = \frac{\cos^2 \theta^{\circ}_{23} \sin^2 \theta \sin^2 \theta^{\circ}_{12} 
+ \cos^2 \theta \sin^2 \theta^{\circ}_{23} 
- \frac{1}{2} \sin 2 \theta \sin 2 \theta^{\circ}_{23} \sin \theta^{\circ}_{12} \cos \phi }{1 - \sin^2 \theta_{13}} \,. 
\label{eq:ss23B1}
\end{equation}

In case B2, the corresponding `sum rules' are
\begin{equation}
\sin^2 \theta_{12} = 1 - 
\frac{\cos^2\theta^{\circ}_{12} \cos^2\theta^{\circ}_{13}  }{1 - \sin^2\theta_{13}}\,,
\label{eq:ss12B2}
\end{equation}
\begin{equation}
\cos \delta = 
\frac{\cos^2 \theta_{13} (\sin^2 \theta^{\circ}_{12} - \cos^2 \theta_{23}) + \cos^2 \theta^{\circ}_{12} \cos^2 \theta^{\circ}_{13} ( \cos^2 \theta_{23} - \sin^2 \theta_{13} \sin^2 \theta_{23} )}
{ \sin 2 \theta_{23} \sin \theta_{13} | \cos \theta^{\circ}_{12} \cos \theta^{\circ}_{13}| (\cos^2 \theta_{13} - \cos^2 \theta^{\circ}_{12} \cos^2 \theta^{\circ}_{13} )^{\frac{1}{2}}}\,,
\label{eq:cosdeltaB2}
\end{equation}
where the expressions for 
$\sin^2\theta_{13}$ and $\sin^2\theta_{23}$ take the following form 
\cite{Girardi:2015rwa} (see also \cite{Blennow:2020snb}):
\begin{equation}
\sin^2 \theta_{13} = \cos^2 \theta^{\circ}_{13} \sin^2 \theta^{\circ}_{12} \sin^2 \theta + \sin^2 \theta^{\circ}_{13} \cos^2 \theta 
+ \frac{1}{2} \sin 2 \theta \sin 2 \theta^{\circ}_{13} \sin \theta^{\circ}_{12} \cos \phi \,, 
\label{eq:ss13B2}
\end{equation}
\begin{equation}
\sin^2 \theta_{23} = \frac{\cos^2 \theta^{\circ}_{12} \sin^2 \theta}{1 - \sin^2 \theta_{13}} \,. 
\label{eq:ss23B2}
\end{equation}

For $G_e = Z_2$ and $G_{\nu} = Z_2$ (pattern C), there are 
two cases, C5 and C9, featuring a sum rule for $\sin^2\theta_{12}$. 
In case C5,
\begin{equation}
\sin^2 \theta_{12} = \frac{\sin^2 \theta^{\circ}_{12}}{1 - \sin^2 \theta_{13}}\,,
\end{equation}
whereas in case C9,
\begin{equation}
\sin^2 \theta_{12} = \frac{\sin^2 \theta^\circ_{12} - \sin^2 \theta_{13}}{1 - \sin^2 \theta_{13}}\,.
\label{eq:ss12C9}
\end{equation}
In both these cases, $\cos\delta$ is unconstrained.

The residual symmetries are assumed to originate from 
a flavour symmetry that holds at some high-energy scale.
The observed lepton mixing pattern of two large and one small 
mixing angles points to non-Abelian finite (discrete) groups. 
The small (in terms of the number of elements) groups 
featuring a 3-dimensional irreducible representation, 
used to unify three lepton families at high energy, are 
$A_4$, $S_4$ and $A_5$.

In~\cite{Girardi:2015rwa}, a comprehensive analysis 
for these flavour symmetry groups 
has been performed, and the values of the fixed parameters 
$\theta^\circ_{ij}$ have been derived. 
In Table~\ref{tab:cases}, we summarise them along with the 
corresponding predictions for $\sin^2\theta_{12}$ and $\cos\delta$, 
for the cases of interest (B1, B2, C5 and C9).
\begin{table}[t]
\small
\centering
\renewcommand*{\arraystretch}{1.5}
\begin{tabular}{|l|ccc|ccc|}
\hline
Case & $\sin^2\theta^{\circ}_{12}$ & $\sin^2\theta^\circ_{23}$ & $\sin^2\theta^\circ_{13}$ & $\sin^2 \theta_{12}$ & $\cos \delta$ (NO) & $\cos\delta$ (IO) \\
\hline
B1 & 1/3 & 1/2 & $-$ & 0.341 & \phantom{$-$}0.278 & $-0.462$  \\
B1A$_5$ & 0.276 & 1/2 & $-$ & 0.283 & \phantom{$-$}0.319 & $-0.531$ \\
B2S$_4$ & 1/6 & $-$ & 1/5 & 0.318 & $-0.131$ & \phantom{$-$}0.219  \\
B2A$_5$ & 0.095 & $-$ & 0.276 & 0.331 & $-1^\ast\phantom{~~~~}$ & $-0.879$ \\
B2A$_5$II & 1/4 & $-$ & 0.127 & 0.331 & \phantom{$-$}0.970 & $\phantom{-}1^\ast\phantom{~~~~}$ \\
C9A$_5$ & 0.345 & $-$ & $-$ & 0.331 & $-$ & $-$ \\
\hline
\end{tabular}
\caption{Cases derived from $A_4$, $S_4$ and $A_5$ 
and predicting $\sin^2\theta_{12}$ within its $3\sigma$ range 
obtained by NuFIT. 
Case B1 is common to all the three groups.  
The values of $\sin^2 \theta_{12}$ ($\cos \delta$)
are obtained using the best-fit values of 
$\sin^2 \theta_{13}$ ($\sin^2 \theta_{13}$ and $\sin^2 \theta_{23}$ 
for NO and IO). 
An asterisk means that physical values of $\cos \delta$  
cannot be obtained employing the best-fit values of the mixing angles.}
\label{tab:cases}
\end{table}
The irrational values of $\sin^2\theta^\circ_{ij}$ appearing for the cases 
originating from $A_5$ symmetry
are related to the golden ratio $r=(1+\sqrt{5})/2$ 
as follows:  
$1/(2 + r) \approx 0.276$,
$1/(4 r^2) \approx 0.095$, 
and $(3 - r)/4 \approx 0.345$~\cite{Girardi:2015rwa}.
The quoted predictions for $\sin^2\theta_{12}$ and $\cos\delta$
are obtained using the best-fit values of $\sin^2\theta_{13}$ and 
$\sin^2\theta_{23}$ (with Super-Kamiokande atmospheric data) 
from NuFIT 6.0~\cite{Esteban:2024eli,NuFITv6Sep2024}, 
see Table~\ref{tab:NuFIT6.0}. 
Case C5 characterised by $\sin^2\theta^\circ_{12} = 1/4$ 
and common to $S_4$ and $A_5$ 
leads to $\sin^2\theta_{12} \approx 0.256$, 
a value excluded at $4.9\sigma$ by NuFIT 
and at $6.1\sigma$ by the first JUNO measurement. 
This is why this case does not appear in Table~\ref{tab:cases}. 
The mixing patterns in Table~\ref{tab:cases} have been studied 
in the context of current and future neutrino oscillation 
experiments in~\cite{Petcov:2018snn,Blennow:2020snb,Blennow:2020ncm}.

We note that the sum rules for $\sin^2\theta_{12}$ and $\cos\delta$ in case 
B1 correspond to those 
of trimaximal (TM) mixing pattern 2 (TM$_2$)~\cite{Grimus:2008tt}:
\begin{equation}
\sin^2\theta_{12} = \frac{1}{3(1-\sin^2\theta_{13})} > \frac13 
\qquad \text{and} \qquad
\cos\delta = \frac{\left(1-2\sin^2\theta_{13}\right) \cot2\theta_{23}}{\sin\theta_{13} \sqrt{2 - 3\sin^2\theta_{13}}}\,,
\end{equation}
and the sum rules for $\sin^2\theta_{12}$ and $\cos\delta$ in case B2S$_4$ 
to those of TM$_1$~\cite{Albright:2008rp}:
\begin{equation}
\sin^2\theta_{12} = \frac{1-3\sin^2\theta_{13}}{3(1-\sin^2\theta_{13})} < \frac13
\qquad \text{and} \qquad
\cos\delta = -\frac{\left(1-5\sin^2\theta_{13}\right) \cot2\theta_{23}}{2\sqrt{2} \sin\theta_{13} \sqrt{1 - 3\sin^2\theta_{13}}}\,.
\end{equation}
TM mixing patterns have been very recently reconsidered in light of the first JUNO measurement~\cite{Zhang:2025jnn,Jiang:2025hvq}.

We would like to mention that there exist other 
phenomenological sum rules for $\cos\delta$. 
The PMNS matrix $U_\mathrm{PMNS} = U_e^\dagger U_\nu$, 
with the unitary matrix $U_\nu$ ($U_e$) originating from the 
diagonalisation of the neutrino (charged lepton) mass term.
Assuming that (i) $U_\nu$ has one of the five symmetry forms, namely, 
tri-bimaximal~\cite{Harrison:2002er,Harrison:2002kp,Xing:2002sw,He:2003rm}, 
bimaximal~\cite{Vissani:1997pa,Barger:1998ta,Baltz:1998ey},%
\footnote{Bimaximal mixing can be obtained, in particular,
from $U(1)$ symmetry associated with the conservation 
of the non-standard lepton charge 
$L^\prime = L_e - L_\mu - L_\tau$~\cite{Petcov:1982ya}.} 
golden ratio type~A~\cite{Kajiyama:2007gx,Everett:2008et}, 
golden ratio type B~\cite{Rodejohann:2008ir,Adulpravitchai:2009bg}, 
or hexagonal~\cite{Albright:2010ap,Kim:2010zub} mixing, 
and (ii)  $U_e$ is given by either a unitary rotation in one plane 
or a product of two unitary rotations in any two different planes,     
sum rules for $\cos\delta$ were derived in~\cite{Petcov:2014laa},
and the respective predictions for $\cos\delta$ 
were analysed in detail in \cite{Girardi:2015vha}.
The test of these phenomenologically derived sum rules 
might provide independent evidence, if any, 
for a non-Abelian discrete symmetry at the origin of the 
experimentally observed peculiar lepton mixing pattern of two large 
and one small mixing angles.

\paragraph{Statistical analysis.}
To assess the viability of the mixing patterns in Table~\ref{tab:cases}, 
we construct an approximate likelihood function, similarly to the procedure discussed in~\cite{Petcov:2018snn}. Namely, for the observable $\alpha$ of interest ($\sin^2\theta_{12}$, $\sin^2\theta_{23}$ or $\cos\delta$), it is given by
\begin{equation}
\mathcal{L}(\alpha) = \exp\left(-\frac{\chi^2(\alpha)}{2}\right)\,,
\label{eq:likelihood}
\end{equation}
where
\begin{equation}
\chi^2(\alpha) = \min \Bigg[\chi^2(\vec{x}) 
\bigg|_{\stackon[2pt]{\small sum rules}{\small $\alpha = \mathrm{const}$}}
\Bigg]\,, 
\label{eq:chi2alpha}
\end{equation}
with
\begin{equation}
\vec{x} = (\sin^2 \theta_{12},\sin^2 \theta_{13},\sin^2 \theta_{23},\delta)\,,
\label{eq:vecx}
\end{equation}
where the theoretical relations between the standard mixing parameters
$\vec{x}$ and the model parameters $(\theta,\phi)$, 
eqs.~\eqref{eq:ss12B1}--\eqref{eq:ss23B1} for patterns B1, 
eqs.~\eqref{eq:ss12B2}--\eqref{eq:ss23B2} for patterns B2, etc., 
have been taken into account.
The NuFIT collaboration does not provide 
the total $\chi^2(\vec{x})$ obtained in the global analysis. 
However, they do provide one-dimensional projections 
$\chi^2_i(x_i)$ for each mixing parameter $x_i$, 
as well as two-dimensional projections 
$\chi^2_{ij}(x_i,x_j)$ for each pair of the mixing parameters
$(x_i,x_j)$, $i \neq j$, 
on their website~\cite{NuFITv6Sep2024}.
We approximate the 
total $\chi^2(\vec{x})$ as the following sum:%
\footnote{We have checked that the approximate $\chi^2(\vec{x})$ 
constructed from the four one-dimensional projections as 
$\chi^2(\vec{x}) = \sum_{i=1}^4 \chi^2_i(x_i)$
reproduces extremely well the two-dimensional 
allowed regions in the planes $(x_i,x_j)$, $i \neq j$,
obtained by NuFIT~\cite{Esteban:2024eli,NuFITv6Sep2024},  
with the only exception being the regions in the plane
$(\sin^2\theta_{23},\delta)$ 
for NO, that are somewhat overestimated. For this reason, we adopt 
the more accurate approximation in eq.~\eqref{eq:chisq}.}
\begin{equation}
 \chi^2(\vec{x}) = \chi^2_1(\sin^2\theta_{12}) + \chi^2_2(\sin^2\theta_{13}) 
 + \chi^2_{34}(\sin^2\theta_{23},\delta)\,.
 \label{eq:chisq}
\end{equation}
In this way, we account with a relatively high precision
for non-trivial correlations between $\sin^2\theta_{23}$ 
and $\delta$ present in the current experimental data,
all other correlations between the observables of interest
being negligible.%
\footnote{This was confirmed also in the discussions with
I.~Esteban, M.~C.~Gonzalez-Garcia and A.~Marrone, who performed
a detailed numerical study of the correlations and provided us with the 
results.}

First, we perform the statistical analysis using the results 
of the global fit to neutrino oscillation data.
Next, to incorporate the JUNO result, we replace $\chi^2_1(\sin^2\theta_{12})$ in eq.~\eqref{eq:chisq} with
\begin{equation}
\chi_\mathrm{1,\,JUNO}^2\left(\sin^2\theta_{12}\right) = \left(
\frac{\sin^2\theta_{12} - 0.3092}{0.0087}
\right)^2\,,
\label{eq:chi2ss12JUNO}
\end{equation}
employing the best-fit value of $\sin^2\theta_{12}$ and 
the corresponding $1\sigma$ error from eq.~\eqref{eq:JUNO}. 
For comparison, the latest NuFIT result is 
$\sin^2\theta_{12} = 0.308^{+0.012}_{-0.011}$, i.e., 
the $1\sigma$ error on $\sin^2\theta_{12}$ reported by JUNO 
is a factor of 1.3 smaller than that derived in the global analysis.

\begin{figure}[t!]
\centering
\includegraphics[width=\textwidth]{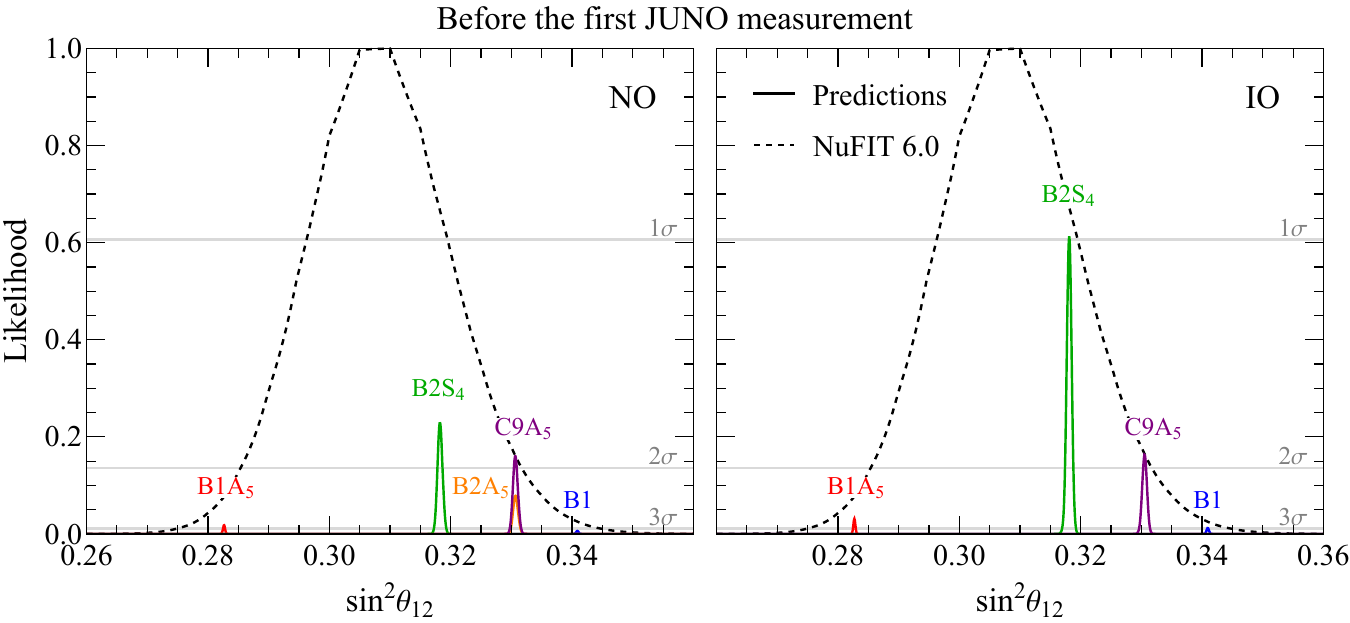}\\[0.1cm]
\includegraphics[width=\textwidth]{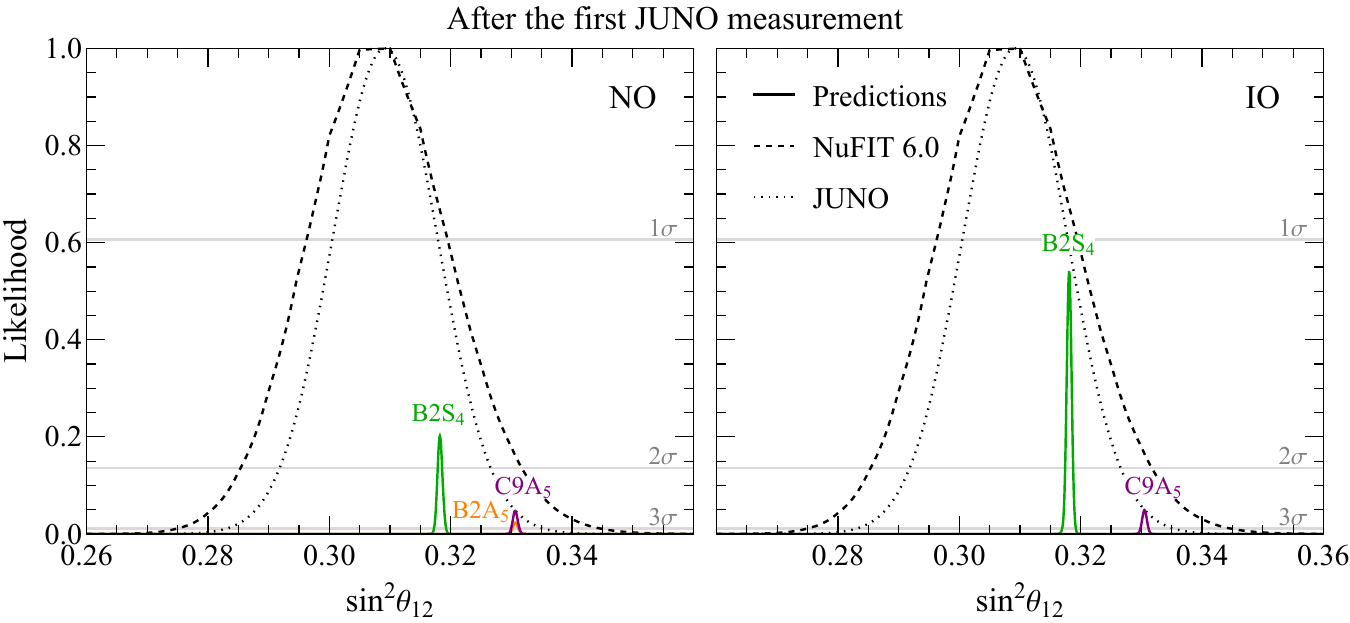}
\caption{\textit{Upper panels:} 
predictions for $\sin^2\theta_{12}$ obtained using the results 
of the global analysis of neutrino oscillation data performed 
by the NuFIT collaboration in September 2024~\cite{Esteban:2024eli,NuFITv6Sep2024}.
\textit{Lower panels:} 
the same after including the first JUNO measurement from November 2025~\cite{JUNO:2025gmd}. 
The black dashed line corresponds to the one-dimensional projection 
$\chi^2_1(\sin^2\theta_{12})$ obtained by NuFIT, whereas 
the black dotted line corresponds to 
$\chi^2_\mathrm{1,\,JUNO}(\sin^2\theta_{12})$ in eq.~\eqref{eq:chi2ss12JUNO}.}
\label{fig:ss12}
\end{figure}

\paragraph{Results.}
In the upper panels of Fig.~\ref{fig:ss12}, we present the likelihood functions 
for $\sin^2\theta_{12}$ for the mixing patterns from Table~\ref{tab:cases}.
The top-left (top-right) panel uses as an input the 
NuFIT one-dimensional projections  
$\chi^2_1(\sin^2\theta_{12})$, $\chi^2_2(\sin^2\theta_{13})$ 
and two-dimensional projection $\chi^2_{34}(\sin^2\theta_{23},\delta)$ 
for NO (IO). 
The global analysis \cite{Esteban:2024eli,NuFITv6Sep2024}
 (with Super-Kamiokande atmospheric data)
finds an overall preference for NO at the level 
of~$2.5\sigma$ ($\D\chi^2 = 6.1$). 
However, we take a conservative approach and treat both NO and IO 
on equal grounds.%
\footnote{This is the reason why the black dashed line reaches 
$\mathcal{L}(\sin^2\theta_{12})=1$ in the maximum
(corresponding to $\chi^2_1(\sin^2\theta_{12}) = 0$)  
for both neutrino mass orderings.}
We note that the main differences between the NuFIT 6.0 results and those 
of NuFIT 3.2 used in~\cite{Petcov:2018snn} are the best-fit values of 
and the one- and two-dimensional $\chi^2$-projections for $\sin^2\theta_{23}$ 
and~$\delta$. As a consequence, 
case B2A$_5$ that was found to be compatible 
with the global data at $3\sigma$ in~\cite{Petcov:2018snn}
is now disfavoured at $3.2\sigma$ for IO, 
and thus, does not appear in the right panel. 
Further, cases B1 and B1A$_5$ barely surpass the $3\sigma$ line. 
Case B2A$_5$II is disfavoured at $3\sigma$ ($4\sigma$) for NO (IO), 
and thus, does not appear in the plots.

In the lower panels of Fig.~\ref{fig:ss12}, 
we implement the JUNO result 
in the form of eq.~\eqref{eq:chi2ss12JUNO}. 
The high-precision measurement of $\sin^2\theta_{12}$ 
results in disfavouring cases B1 (corresponding to TM$_2$) 
and B1A$_5$ at $3.8\sigma$ and $3.3\sigma$, 
respectively, for both NO and IO. 
In addition, case B2A$_5$,  
while being compatible with the JUNO result at 
$2.7\sigma$ for NO,
is disfavoured at $3.5\sigma$ for IO. 
Overall, out of the five (four) cases 
allowed by the global data at $3\sigma$ for NO (IO), 
only three (two) cases survive at the same confidence level 
after the JUNO measurement. 
The most favoured pattern, compatible with the data at $1.8\sigma$ ($1.1\sigma$) 
for NO (IO) is case B2S$_4$ corresponding to TM$_1$.

\begin{figure}[t!]
\centering
\includegraphics[width=\textwidth]{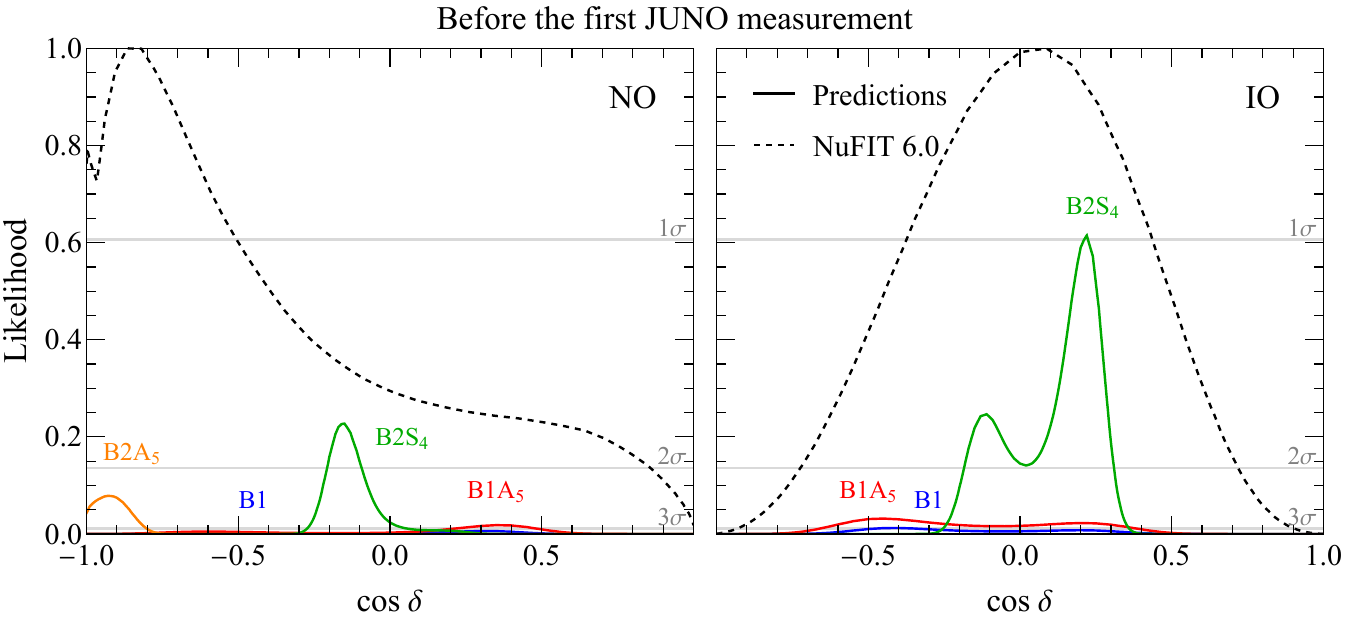}\\[0.1cm]
\includegraphics[width=\textwidth]{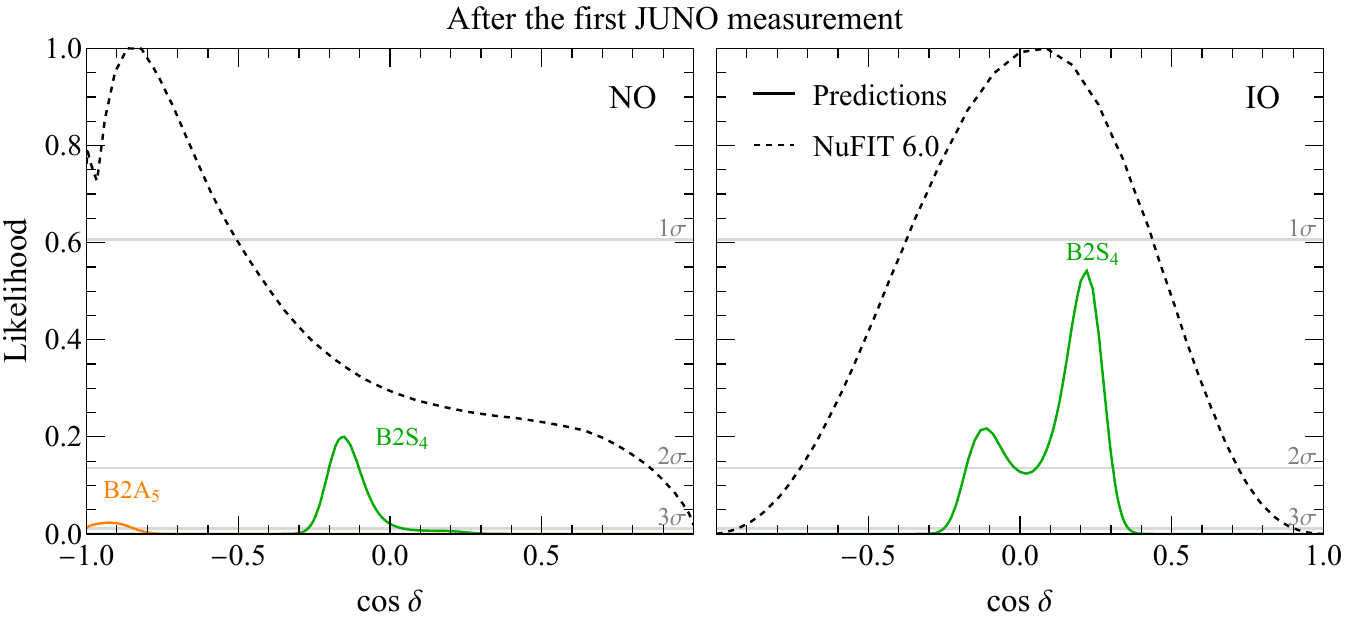}
\caption{\textit{Upper panels:}
predictions for $\cos\delta$ obtained using the results 
of the global analysis of neutrino oscillation data performed 
by the NuFIT collaboration in September 2024~\cite{Esteban:2024eli,NuFITv6Sep2024}. 
\textit{Lower panels:}
the same after including 
the first JUNO measurement from November 2025~\cite{JUNO:2025gmd}.
The black dashed line corresponds to the one-dimensional projection 
$\chi^2_4(\delta)$ obtained by NuFIT.}
\label{fig:delta}
\end{figure}

In Fig.~\ref{fig:delta}, we show the predictions of patterns B for $\cos\delta$ 
before (upper panels) and after (lower panels) the JUNO result.
For these patterns, the predicted value of $\cos\delta$ is very sensitive 
to the value of $\theta_{23}$, as can be understood 
from eqs.~\eqref{eq:cosdeltaB1} and \eqref{eq:cosdeltaB2}.
The two maxima in $\mathcal{L}(\cos\delta)$ for IO, 
especially pronounced in case B2S$_4$, 
are a consequence of the fact 
that the one-dimensional projection $\chi^2_3(\sin^2\theta_{23})$ has two local minima, 
unlike in the NuFIT 3.2 results used to obtain Fig.~3 in~\cite{Petcov:2018snn}.
As we can see from the lower panels, the JUNO measurement 
almost does not affect the width of the likelihood profiles, 
that is driven by the uncertainty in the determination of the atmospheric mixing angle, $\theta_{23}$. 
Instead, it has an impact on the overall compatibility of the mixing patterns 
with the data, with only cases B2S$_4$ and B2A$_5$ (B2S$_4$) 
surviving at the $3\sigma$ confidence level for NO (IO). 
We recall that case C9A$_5$ does not yield a sum rule relating $\cos\delta$ 
to the mixing angles, and hence, it is not present in Fig.~\ref{fig:delta}.

\begin{figure}[t!]
\centering
\includegraphics[width=\textwidth]{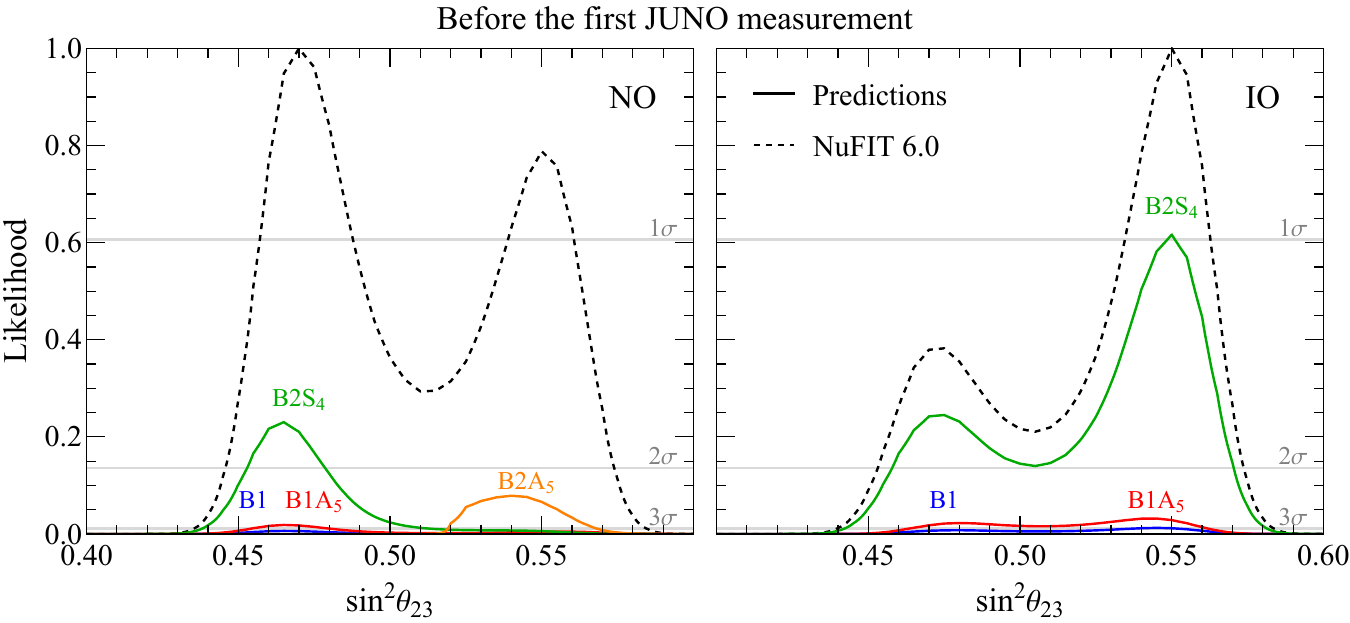}\\[0.1cm]
\includegraphics[width=\textwidth]{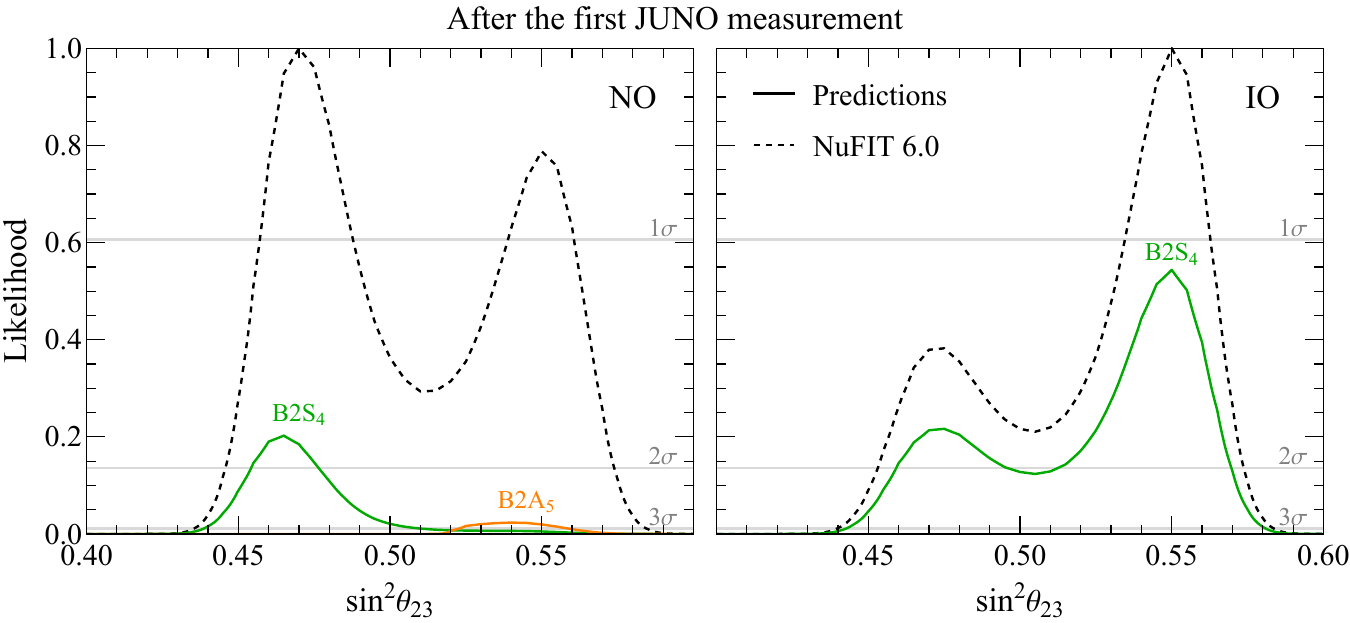}
\caption{\textit{Upper panels:}
predictions for $\sin^2\theta_{23}$ obtained using the results 
of the global analysis of neutrino oscillation data performed 
by the NuFIT collaboration in September 2024~\cite{Esteban:2024eli,NuFITv6Sep2024}. 
\textit{Lower panels:}
the same after including 
the first JUNO measurement from November 2025~\cite{JUNO:2025gmd}.
The black dashed line corresponds to the one-dimensional projection 
$\chi^2_3(\sin^2\theta_{23})$ obtained by NuFIT.}
\label{fig:ss23}
\end{figure}

For completeness, in Fig.~\ref{fig:ss23} we show the predictions of patterns B for $\sin^2\theta_{23}$ 
before (upper panels) and after (lower panels) the JUNO result.
The corresponding likelihood profiles have been obtained 
by solving eq.~\eqref{eq:ss23B1} (eq.~\eqref{eq:ss23B2}) in cases B1 (B2)
for $\theta$ for a given $\phi \in [0,2\pi)$, 
or alternatively, by solving the equation for $\phi$ 
for a given $\theta \in [0,\pi)$. 
In case B2S$_4$ favoured by the data, the shape of the likelihood for IO
follows that of the likelihood derived in the global analysis (black dashed line), while its magnitude is suppressed, 
mainly because of the sharp prediction for $\sin^2\theta_{12}$. 
Almost all the values of $\sin^2\theta_{23}$ from its experimental $3\sigma$ 
range are allowed at $3\sigma$ for IO, 
while $\sin^2\theta_{23} \in [0.438,0.513]$ at $3\sigma$ for NO.
In case B2A$_5$,  
the values of $\sin^2\theta_{23} \in [0.519,0.568]$ 
are compatible with the global data at $3\sigma$ assuming NO. 
Taking into account the JUNO result, the allowed range is slightly reduced to 
$\sin^2\theta_{23} \in [0.523,0.559]$. 
As we have already seen, case B2A$_5$ is disfavoured at more than $3\sigma$ 
assuming IO. 
The likelihood profiles in cases B1 and B1A$_5$ are relatively broad, 
especially for IO. 
However, both cases are disfavoured at almost $3\sigma$ 
by the global data, 
and at more than $3\sigma$ after taking into account the JUNO measurement.

Since any of the non-Abelian symmetries discussed by us 
are assumed to be realised at some high scale, 
the question arises how our results might be modified 
by renormalisation group (RG) running effects. 
The evaluation of the RG effects requires additional specific assumptions 
about the scale at which the symmetry is realised, 
the underlying model (Standard Model (SM), minimal supersymmetric SM, etc.), 
the nature~---~Dirac or Majorana~---~of massive neutrinos 
and the mechanism of neutrino mass generation 
(SM minimally extended with right-handed singlet neutrinos 
and conserved total lepton charge for Dirac neutrinos, 
or seesaw mechanism/Weinberg dimension-5 operator 
for Majorana neutrinos, etc.). 
They further depend on the type of spectrum of neutrino masses (NO or IO), 
on the value of the lightest neutrino mass and, 
in the case of massive Majorana neutrinos, on the Majorana phases. 
However, the RG effects are known to be negligible 
for the NO mass spectrum if the lightest neutrino mass is smaller 
than approximately 
0.01~eV; for the IO spectrum and massive Majorana neutrinos, 
the RG effects on $\theta_{12}$ can be large, 
but are known to be particularly strongly suppressed 
for, e.g., specific values of the Majorana phases~\cite{Antusch:2003kp} 
(see also, e.g.,~\cite{Petcov:2005yh}).
Thus, we can conclude conservatively that our results 
are valid at least in the indicated  cases. 
For detailed studies of the RG effects on sum rules 
arising for various lepton mixing patterns, 
we refer the reader to~\cite{Zhang:2016djh,Gehrlein:2016fms,Zhang:2025jnn}.

\paragraph{Summary and conclusions.}
In this \textit{letter}, we studied the implications of the first JUNO measurement of $\sin^2\theta_{12}$~\cite{JUNO:2025gmd} 
for lepton mixing patterns shaped by residual symmetries 
of the charged lepton and neutrino mass matrices, that originate from 
$A_4$, $S_4$ and $A_5$ flavour symmetries. 
First, we updated our previous analysis~\cite{Petcov:2018snn} of these mixing patterns 
using the results of the latest global fit to neutrino oscillation data 
performed by NuFIT~\cite{Esteban:2024eli,NuFITv6Sep2024}. 
We found that among the mixing patterns yielding sharp predictions for $\sin^2\theta_{12}$, 
there are five (four) cases compatible with the global data at 
$3\sigma$ for NO (IO) (upper panels of Fig.~\ref{fig:ss12}). 
Next, we performed a statistical analysis incorporating the recent JUNO result 
in eq.~\eqref{eq:JUNO} and found that the number of cases viable at $3\sigma$ 
is reduced to only three (two) for NO (IO)
(lower panels of Fig.~\ref{fig:ss12}).
We further derived the likelihood functions for $\cos\delta$ 
and $\sin^2\theta_{23}$ before and after the JUNO measurement, see Figs.~\ref{fig:delta} and \ref{fig:ss23}, respectively. 
Our results show that  the first JUNO data 
has strong implications for the lepton mixing patterns
arising from $A_4$, $S_4$ and $A_5$ flavour symmetries. 
It is very probable that all of the remaining viable cases will be 
critically tested by future JUNO measurements of $\sin^2\theta_{12}$.
However, if any of them survives the future tests, 
plausibly this could be case B2S$_4$ corresponding to TM$_1$ mixing,
a sufficiently high-precision measurement of $\delta$ 
would provide a decisive answer on whether it is realised in nature.

\small
\paragraph{Acknowledgements.}
We thank  I.~Esteban, M.~C.~Gonzalez-Garcia and A.~Marrone 
for useful discussions on the statistical analysis.
A.V.T. is funded by the European Union, NextGenerationEU, 
National Recovery and Resilience Plan
(mission~4, component~2) 
under the project \textit{MODIPAC: Modular Invariance in Particle Physics and Cosmology} (CUP~C93C24004940006). 
A.V.T. is further supported by the INFN Iniziativa Specifica 
\textit{APINE: Physics at the Energy, Intensity, and Astroparticle Frontiers}.
The work of S.T.P. was supported in part by 
the European Union’s Horizon Europe research and innovation programme 
under the Marie Sk\l{}odowska-Curie Staff Exchange grant agreement No.~101086085-ASYMMETRY,  
by the Italian INFN program on Theoretical Astroparticle Physics and by 
the World Premier International Research Center Initiative 
(WPI Initiative, MEXT), Japan.

\bibliography{ViabilityA4S4A5JUNO2025}

\end{document}